\documentclass[11pt,twoside]{article}
\usepackage{asp2010}

\resetcounters

\begin{document}

\title{A Swiss Watch Running on Chilean Time:  A Progress Report on Two New Automated CORALIE RV Pipelines}
\author{J.S.~Jenkins$^1$ and A.~Jord\'an$^2$
\affil{$^1$Departamento de Astronom\`ia, Universidad de Chile, Camino el Observatorio 1515, Las Condes, Santiago, Chile Casilla 36-D}
\affil{$^2$Departamento de Astronom\`ia y Astrof\`isica, Pontificia Universidad Cat\`olica de Chile, 7820436 Macul, Santiago, Chile}}

\begin{abstract}
We present the current status of two new fully automated reduction and analysis pipelines, built for the Euler
telescope and the CORALIE spectrograph.  Both pipelines have been designed and built independently at the
Universidad de Chile and Universidad Catolica by the two authors.  Each pipeline has also been written on two
different platforms, IDL and Python, and both can run fully automatically through full reduction and analysis of
CORALIE datasets.  The reduction goes through all standard steps from bias subtraction, flat-fielding, scattered
light removal, optimal extraction and full wavelength calibration of the data using well exposed ThAr arc lamps.
The reduced data are then cross-correlated with a binary template matched to the spectral type of each star and
the cross-correlation functions are fit with a Gaussian to extract precision radial-velocities.  For error analysis we are currently testing
bootstrap, jackknifing and cross validation methods to properly determine uncertainties directly from the data.
Our pipelines currently show long term stability at the 12-15m/s level, measured by observations of two known
radial-velocity standard stars.  In the near future we plan to get the stability down to the 5-6m/s level and also
transfer these pipelines to other instruments like HARPS.
\end{abstract}

\section{Introduction}

The CORALIE spectrograph is a proven instrument for the detection of extrasolar planets (aka exoplanets) around stars like 
the Sun via the radial-velocity (RV) method, with a number of exciting discoveries already having been made using this instrument 
(e.g. \citealp{queloz00}; \citealp{eggenberger06}; \citealp{segransan10}).  Hunting for exoplanets with CORALIE requires knowledge and application of the so called \emph{simultaneous Thorium} 
technique, which does not employ any absorption cell in the light beam but instead is based around the concept of point spread function stablisation 
throughout the optical train of the instrument.  Two fibres are used; fibre A monitors the star under observation and feeds the light to the 
spectrograph, whereas fibre B simultaneously feeds light from a Thorium-Argon (ThAr) gas lamp to the spectrograph, such that the ThAr 
lines can serve as a reference to monitor the drift of the spectrograph throughout the observation \citep[see][]{baranne96}.

CORALIE has been used as the test bed for the successful ESO-HARPS spectrograph \citep{pepe00} but in contrast to HARPS it is mounted on a telescope 
with a primary mirror diameter of only 1.2m, it operates at around half the resolving power of HARPS and it does not actively stabilise environmental 
pressure changes that will cause the RV zero point to drift throughout an observing night due to changes in the refractive index of air.  In spite of this, 
and as highlighted above, CORALIE is more than sufficient to discover and fully characterise planetary mass bodies with various orbital characteristics 
orbiting bright solar-type stars, particularly since the June 2007 hardware upgrade of the instrument that gives rise to a gain in magnitude of 1.5.

We have began a number of RV projects that can be accomplished using CORALIE.  One such project is part of the Calan-Hertfordshire Extrasolar Planet Search 
(CHEPS) which is a search for new short period gas giant planets that have a high probability of transiting their host star (see \citealp{jenkins08}; \citealp{jenkins09}).  
Another CORALIE project shall be the follow-up of transiting systems detected by the new HAT-South project \citep{bakos09}.

\section{Pipeline Steps}

Here we provide a rundown of the steps we are employing in our pipelines to measure the precision RVs necessary to detect and characterise planets orbiting 
other stars.  We note that both Jenkins and Jord\'an are independently developing separate pipelines for CORALIE, with the Jenkins pipeline being constructed primarily 
in IDL and the Jord\'an pipeline being built in Python.

\subsection{Reduction and Extraction}

\begin{figure}[!ht]
\plotone{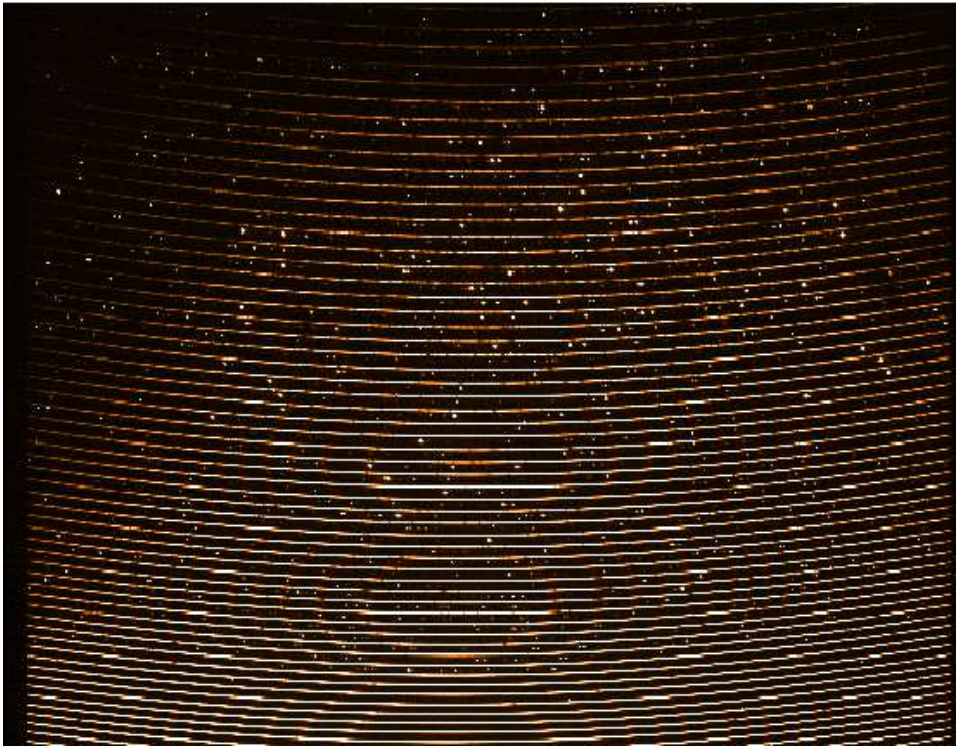}
\caption{CORALIE spectral formal.  All continuous curved orders represent the science orders from fibre A and the fainter orders parallel to these 
that exhibit bright spots (ThAr emission features) are the calibration orders from fibre B.}
\end{figure}

The reduction steps follow normal methods for properly reducing and extracting high resolution echelle spectra.   First, the bias
signal is removed from each individual frame by making use of the biases obtained as part of the CORALIE standard calibration plan.  
The residual overscan region is then subtracted out and trimmed off the image.  The CORALIE spectral format is shown in Fig.~1 and consists 
of 72 spectral orders for fibre A (the object fibre) and a further 50 orders for fibre B (ThAr fibre).  These orders are precisely
traced using a well exposed flatfield image to properly trace the light path of the order in the dispersion direction.  The traces
are then sigma-clipped to remove any stray pixel counts and to help tighten up the trace and then the object and background apertures are
defined.  To correct for the CCD pixel-to-pixel response we use a number of well exposed flatfields with counts above 10000 that we median
combine together to create a master flatfield frame.  The master flats are then normalised and used to flatfield each image and then the 
scattered light is removed.  This is done by setting the dekker limits large enough 
to include wide inter-order spacing regions and a low-order polynomial is used to sample any gradients along the orders.  The profile of each order 
is then measured by sub-sampling each order profile individually using Gaussians.  This model is finally used to extract the object and ThAr orders using 
an optimal extraction algorithm \citep{horne86}.  Finally the extracted master flatfield is then divided into all extracted science orders to remove any 
low frequency residuals not removed by the initial flatfield procedure and also to correct for the instrumental blaze function.

\subsection{Post-extraction RV Analysis}

\subsubsection{Wavelength Calibration}

The first and crucial step after the extraction of all spectra is to properly determine the wavelength solution for each order.  In contrast to the 
absorption cell method, we have no embedded absorption lines to serve as an exact wavelength fiducial directly over the stellar spectra, therefore 
we require a high level of precision and consistency in all our ThAr wavelength solutions.  \citet{lovis07} have derived a new list of absorption lines 
using the ThAr gas lamps on HARPS that is anchored to the older and widely used list of lines from \citet{palmer83} and we use a subset of 
lines from this atlas for our wavelength determinations.  

We decided to focus on utilising only the Thorium lines in our wavelength solutions as a number of authors have shown that Argon lines tend to 
be more sensitive to changes in pressure, such as the changes in pressure that differing lamps will operate at, and hence if their energy levels are 
affected by this change, then so will the transition energies between levels be affected, leading to a pressure shift in velocity for Argon lines.  In 
reality we do not find any difference by including Argon lines in the wavelength solution at the current stage of our pipeline development but we 
would expect to in the future once we approach the fundamental CORALIE RV precision limit.

\begin{figure}[!ht]
\plotfiddle{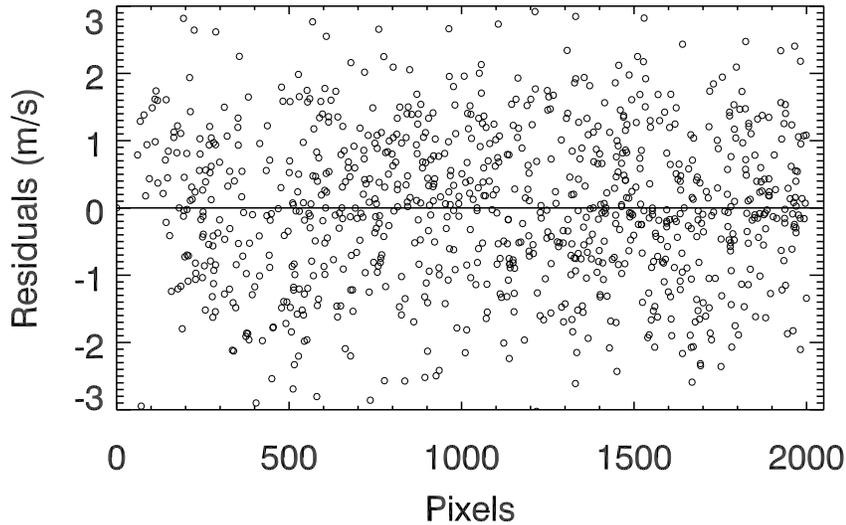}{2.1cm}{90}{50}{50}{185}{-170}
\vspace{4.1cm}
\caption{A sample wavelength solution for a single CORALIE Thorium (fibre B) observation.  The residuals of each individual line against the solution is 
plotted as a function of their pixel position.  Over 900 lines across the whole 50 orders are included and we can see that all points lie within $\pm$3m/s 
from the fit solution.  The rms scatter for this example is 1.15m/s}
\end{figure}

Using the Thorium lines we employ two separate strategies to wavelength calibrate the data.  One method calibrates all orders independently using 
our input of selected lines and the other method employs a global wavelength solution that effectively fits all orders simultaneously using realisations 
of the grating equation.  In both methods we fit the individual lines using gaussians, and for blended lines we use the superposition of two, three or 
four gaussians depending on the blending factor for the region, using non-linear least squares methods.  In the case of the Jenkins pipeline we use 
the MPFIT IDL package developed by \citet{markwardt09}.  We aim to get wavelength solutions with an rms scatter on the zero point better than 80m/s by 
iteratively rejecting outliers and then refitting the solution until we reach our desired level of precision.  Fig.~2 shows 
the residuals for each individual line from an example solution for the calibration fibre B.  There are over 900 lines included and we find an rms to 
this solution of only 1.15m/s, highlighting our robust method for wavelength determination.  Given CORALIE is not in full vacuum isolation in a similar 
manner to HARPS, we observe 
double Thorium calibration frames at least every two hours throughout each night to reset the wavelength zero point and when combined with the 
calibration frames we generally have around 15-20 double Thoriums to wavelength calibrate.

\subsubsection{Cross-Correlation Measurements}

The final part of the pipeline is the measurement of the RV of the star and the velocity drift of the instrument.  We perform a cross-correlation between 
the observed star and a binary template mask that is a close match to the spectral type of the star in question.  The binary masks were developed for use 
on HARPS therefore we actively change the width of the binary emission holes to conform to the resolution of CORALIE \citep{pepe02}.  The cross-correlation 
function (CCF) for each order is then combined using a custom weighting scheme and is fit by a gaussian to measure the systemic velocity.  The drift 
correction is measured in two ways; firstly we build a CCF between the double Thorium calibration and the science fibre B, combine order-by-order, and 
fit again by a gaussian, secondly we also previously calculate directly the wavelength solution from the Thoriums in fibre B for the science frames and 
then determine the mean difference in wavelength using the Doppler equation to translate this into a mean velocity.  The drift correction is then subtracted 
from the RV measured from fibre A and we finally obtain the actual stellar RV.

\section{Standard Stars}

To assess the stability and precision of our new pipelines we have been continuously monitoring two stars that have been observed over the long term and 
have been shown to be RV stable at around the 2m/s level (A. Howard, private communication).  These stars are HD72673 and HD157347.  

HD72673 is classed as a G9V ($B-V$=0.78), high proper motion star and with a parallax of 81.81$\pm$0.46 \citep{floor07}, it is located at distance of only 
12.2pc, giving rise to a visual magnitude of 6.38.  This star has been observed over the long term at Keck and has been shown to be stable at better than 
the 2m/s level.  HD157347 has a spectral type of G5IV ($B-V$=0.68) and with a Hipparcos parallax of 51.22$\pm$0.40, it is located at a distance of 19.5pc.  The close proximity 
and spectral type of the star couple to give a visual magnitude of 6.28, slightly brighter than HD72673.  This star has also been monitored at Keck over the long term 
and has been found to exhibit no significant RV variations down to the 2m/s level.  This highlights that both these stars should serve are good diagnostic 
systems to assess the stability of our pipelines.

\subsection{Current Results}

\begin{figure}[!ht]
\plotfiddle{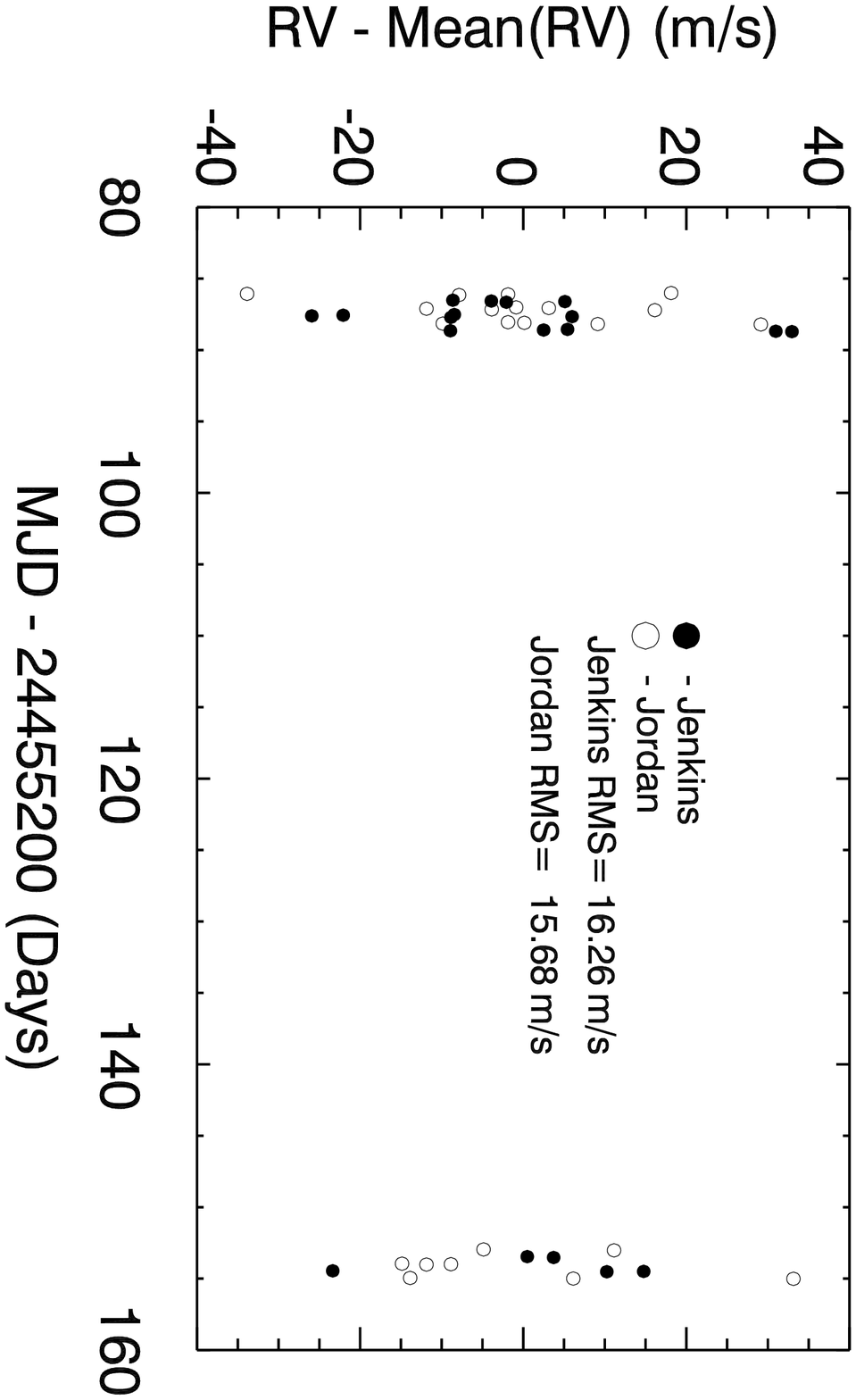}{2cm}{90}{40}{40}{150}{-120}
\plotfiddle{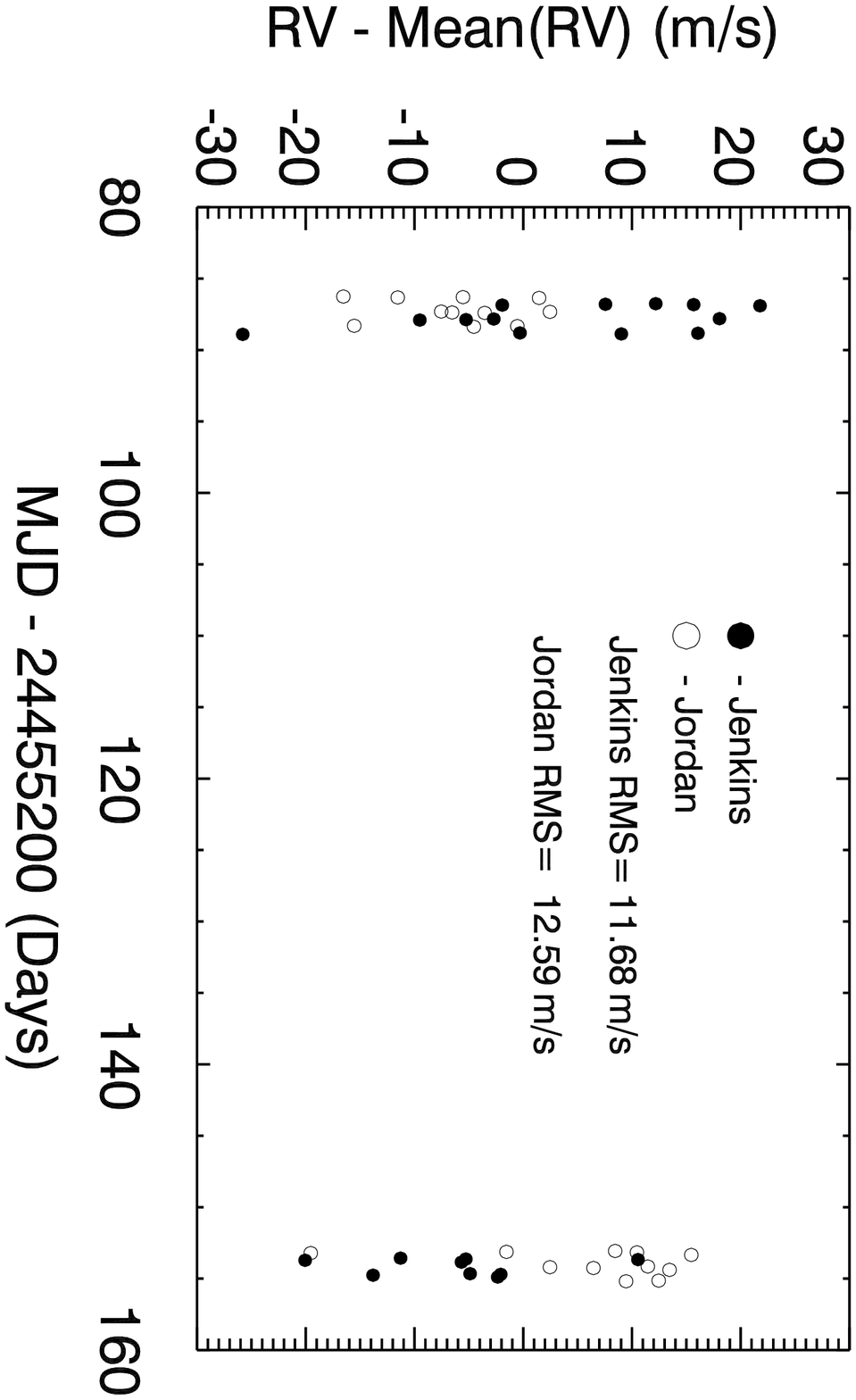}{2cm}{90}{40}{40}{150}{-215}
\vspace{6.3cm}
\caption{CORALIE RVs for the known RV standard stars HD72673 (top panel) and  HD157347 (lower panel).  The filled circles are the data points from the IDL pipeline 
from Jenkins and the open circles are from the Python pipeline of Jord\'an.  The rms scatter around a linear fits are shown in each plot.}
\end{figure}

In Fig.~3 we show all current velocities for our standard stars HD72673 (top) and HD157347 (bottom).  The filled circles are from the pipeline of Jenkins and the open circles 
represent the data from the pipeline of Jord\'an.  In both plots the rms scatter around a linear trend (noise model) is show for both pipelines.  It is clear that both pipelines 
are currently at the same level of long term (few months) precision, which is found to be $\sim$14m/s.  We also note that we find some clustering of data points throughout 
individual nights, so the nightly precision is somewhat higher than the precision we quote here.

14-15m/s long term RV precision was the precision originally quoted in the \citeauthor{baranne96} paper for the ELODIE spectrograph.  This is very similar to what we are finding 
with our pipelines at their current beta-testing level.  Clearly there are a number of improvements that can be made to our pipelines and some of these are discussed below.

\section{Upgrades In Progress}

\subsection{Wavelength Calibration}

One of the most crucial stages in this method is the wavelength calibration for both fibre A and B.  We have implemented a stable scheme that we have outlined above, however 
improvements can be made to our routines and we are in the process of testing these improvements.  Currently we are fitting each Thorium line with a gaussian and employ 
an un-weighted scheme for finding the centroids.  \citet{queloz01} mention that on CORALIE there is an associated uncertainty of $\sim$50m/s that must be accounted for 
due to correctly finding the photo center of each pixel to better determine each lines centroid position.  We are working on a scheme at the moment to better correct for this 
uncertainty and we expect this should lower our long term precision.

\subsection{Uncertainties}

Another important step in any automated RV reduction and analysis pipeline is to properly determine the final uncertainties on any RV data point.  Both on CORALIE and HARPS 
using the simultaneous Thorium reference method, a semi-theoretical approach is employed whereby all the instrumental uncertainties are considered to be well defined. 
\citet{bouchy01} explained the concept behind this method of obtaining the measurement uncertainties based on estimating the quality factor ($Q$) for a given type of stellar 
spectrum.  This $Q$-factor can be thought of as a measure of the amount of spectral information contained in each stellar spectrum i.e. stars with many strong, deep and 
unblended lines have a higher $Q$-factor than rapidly rotating, metal-poor, early type stars that have few lines which are weak, broadened and blended.  This method has 
proven to be fairly robust since it targets areas that most precision RV surveys already consider, such as weeding out stars that are rapidly rotating \citep[see][]{jenkins09b}.

In comparison, we are in the process of testing a more direct approach to measure the uncertainties in our RV determination method, an approach closer to the absorption cell method.  We 
plan to employ a bootstrap method to scramble the CCFs before the weighted combination of each one into the final mean CCF for RV computation.  This scrambling should 
allow us to test how robust the RV zero point is against our weighting scheme.  In addition we plan to test jackknifing and cross validation methods which are similar to the 
bootstrapping scheme except they sample the distribution in a different manner and should allow us to test how sensitive our RVs across our individual orders.

\section{Summary}

We have given a status update on ongoing work to develop two new Doppler pipelines for the CORALIE spectrograph located on the ESO la Silla site in 
Chile.  We have explained the pipeline steps as they currently stand and have highlighted a few steps we are currently working on to increase the stability 
and precision of our codes.  Finally we have shown that the pipelines have thus far reached a precision of around 14m/s, sufficient to discover and characterise 
planets around stars like the Sun.

\acknowledgements JSJ acknowledges funding by Fondecyt through grant 3110004, along with partial support from Centro de Astrof\`\i sica FONDAP 15010003, the  
GEMINI-CONICYT FUND and from the Comit\`e Mixto ESO-GOBIERNO DE CHILE.  AJ acknowledges support from Fondecyt project 1095213, BASAL CATA
PFB-06, FONDAP CFA 15010003 and MIDEPLAN ICM Nucleus P07-021-F.  We would like to acknowledge the helpful discussions with members of the 
Geneva group, in particular Christophe Lovis, Francesco Pepe and Dominique Naef, and also the various support astronomers stationed at the Euler telescope.

\bibliography{jenkins_j}

\end{document}